\documentclass[aps,superscriptaddress,showpacs,nofootinbib,twocolumn]{revtex4}

\usepackage[dvips]{color}  
\usepackage{eucal}

\usepackage{epsfig}

\begin{document}

\title{Finite Temperature Phase Diagram of Quasi-Two-Dimensional Imbalanced Fermi Gases Beyond Mean-Field}
\author{M. A. Resende} \author{A. L. Mota} \author{R. L. S. Farias} \author{Heron Caldas}\email{hcaldas@ufsj.edu.br}
\affiliation{Departamento de Ci\^{e}ncias Naturais, Universidade Federal de S\~{a}o Jo\~{a}o del Rei,\\ 36301-160, S\~{a}o Jo\~{a}o del Rei, Minas Gerais, Brazil}

\begin{abstract}

We investigate the superfluid transition temperature of quasi-two-dimensional imbalanced Fermi gases beyond the mean-field approximation, through the second-order (or induced) interaction effects. For a balanced Fermi system the transition temperature is suppressed by a factor $\approx 2.72$. For imbalanced Fermi systems, the polarization and transition temperature of the tricritical point are significantly reduced as the two-body binding energy $|\epsilon_B|$ increases.

\end{abstract}

\pacs{03.75.Ss, 71.30.+h, 36.20.Kd, 11.10.Kk}

\maketitle

\section{Introduction}

The domain of advanced and accurate experimental techniques of laser cooling and magnetic trapping, as well as imaging of neutral ultracold two-spin-component atomic Fermi gases, has permitted several crucial investigations in these systems. One of the very important studies achieved is the crossover from the Bardeen-Cooper-Schrieffer (BCS) phase of weakly bound Cooper pairs to the Bose-Einstein condensate (BEC) phase of strongly bound diatomic molecules in three-dimensional (3D) trapped Fermi gases~\cite{Exp1,Exp2,Exp3}.

The lowering of dimensionality in many-body quantum systems opens a window for the appearance of interesting phenomena, such as the (yet unexplained) effect of high-temperature superconductivity in two-dimensional (2D) cuprates. Thus, the trap geometry, which is also currently under full control, is fundamental in the use of ultracold atomic gases in the simulation of condensed matter systems, such as layered 2D strongly correlated superconductors. 

As examples of very recent experimental achievements in cold balanced Fermi systems in low dimensions, more specifically in 2D, we highlight the report of the measurement of the density profile and temperature of a 2D gas of atoms~\cite{Turlapov}, pairing in a harmonically trapped 2D atomic Fermi gas in the regime of strong coupling~\cite{Michael1}, and the detection of a many-body pairing gap above the superfluid transition temperature (the pseudogap phenomenon)~\cite{Michael2}. To the best of our knowledge, up to now the only experimental investigations with imbalanced 2D Fermi gases concerns the Fermi polaron problem, in which a single spin-$\downarrow$ atom interacts strongly with a Fermi sea of spin-$\uparrow$ atoms~\cite{nature_khol}.

According to the Mermin-Wagner-Hohenberg-Coleman (MWHC) theorem~\cite{MWHC}, the long-range order is destroyed by fluctuations in uniform, 2D systems. This prohibits the formation of a superconducting phase with a homogeneous order parameter, associated with the breaking of a continuous symmetry. Nevertheless, 2D systems may undergo a phase transition to a state with quasi-long range order via the Berezinskii-Kosterlitz-Thouless (BKT) transition~\cite{BKT}. Since this transition does not require symmetry breaking, it is not forbidden by the MWHC theorem. 

By taking into account the phase fluctuation effect, the BKT transition has been investigated in 2D balanced~\cite{Sa,Duan2} as well as imbalanced systems~\cite{Devreese}. In all these works a strong dependence of the BKT transition temperature $T_{\rm BKT}$ on the two-body energy $|e_B|$ has been found. Below $T_{\rm BKT}$ pairs of vortices and antivortices surge, and will eventually condense as the temperature is lowered~\cite{Sa,Duan2,Devreese}.

The direct consequence of the MWHC theorem is that one should find $T_c=0$ for the phase transition temperature between the normal and superfluid phases in a pure 2D system. However, quasi-2D systems (as, for example, a stack of planes where tunneling between different planes is completely suppressed by a large trapping potential, constituting effective 3D systems) may exhibit a superfluid phase transition at a finite $T_c$ with no conflict with the MWHC theorem~\cite{Lizardo}.

In this paper we study the finite-temperature ($T$) ground state of a quasi-2D atomic Fermi gas with chemical potential imbalance, beyond the mean-field approximation. The standard mean-field calculation does not take into account the effects of the medium on the two-body interaction. This correction, considered first by Gorkov and Melik-Barkhudarov (GMB)~\cite{GMB61}, has been referred to as induced interactions~\cite{Pethick00}, and was found to suppress the mean-field critical temperature of a 3D balanced Fermi gas by a factor $\approx 2.22$~\cite{GMB61,Pethick00,Baranov2,Yu09}. In addition, it has been shown that the GMB correction substantially reduces the order parameter in 2D and 3D lattices~\cite{Torma}. The influence of induced interactions in a 3D imbalanced Fermi gas has been taken into account in Ref.~\cite{Yu10}, and it was found that the polarization $P$ and the transition temperature $T$ of the tricritical point are both reduced from the mean-field results by a factor of about $2.22$, meaning that the transition temperature suppression is comparable with the one found for the spin-balanced configuration.

The possibility of the Fulde-Ferrel-Larkin-Ovchinnikov (FFLO) state with modulated order parameter~\cite{FFLO} is ignored in this work. As in~\cite{GMB61,Pethick00,Baranov,Baranov2,Torma,Yu09}, we consider only pairing between atoms with equal and opposite momenta.

The paper is organized as follows: In Sec.~\ref{TP} we obtain the finite temperature thermodynamic potential of the model in the mean-field approximation. In Sec.~\ref{BMF} we review some basic zero temperature mean-field results which will be used latter. In Sec.~\ref{GMB} we construct the finite temperature phase diagram beyond the mean-field approximation through the GMB correction. We conclude in Sec.~\ref{Conc}.

\section{The Model Hamiltonian and the Mean-Field Thermodynamic Potential}
\label{TP}

We start by considering a 2D nonrelativistic dilute system of fermionic atoms of mass $M$, with two hyperfine states labeled as $\sigma = \uparrow, \downarrow$. This spin-$\uparrow$ and -$\downarrow$ mixture could, in principle, be obtained in 2D experiments with the two lowest hyperfine states of $^{40}{\rm K}$~\cite{Michael1,Michael2}, or with $^{6}{\rm Li}$ atoms, as in the 3D experiments~\cite{Exp5,Exp6,Exp7}.  Their single-particle dispersion relations are given by $\xi_k=\frac{\hbar^2 k^2}{2 M}$. Throughout the paper we set $\hbar=1$. The quasi-2D system can be modeled by the following pairing Hamiltonian:

\begin{equation}
\label{H0}
H=H_{0} + H_{\rm int},
\end{equation}
where

\begin{equation}
\label{H}
H_{0}=\sum_{k,\sigma = \uparrow, \downarrow} \epsilon_k^{\sigma} \psi_{\sigma}^{\dagger}(k) \psi_{\sigma} (k),
\end{equation}
is the kinetic (free) part of $H$ and $H_{\rm int}$ is given below. $\psi_{\sigma}^{\dagger}(k)$ and $\psi_{\sigma}(k)$ in Eq.~(\ref{H}) are the creation and annihilation operators, respectively, for the $\uparrow$ and $ \downarrow$ particles. To assure population imbalance in the system, we have introduced different chemical potential for the species $\sigma$ as $\mu_{\sigma}=\mu + \sigma h$, where $\sigma h  \equiv \pm h$. Then, the chemical potentials $\mu_{\sigma}$ fixes the number densities $n_{\sigma}$ of the different fermions. The new dispersions for the free species $\sigma$, relative to their Fermi energies, are $\epsilon_k^{\sigma} \equiv  \xi_k  -\mu_{\sigma}$. The interaction Hamiltonian is given by

\begin{equation}
H_{int}= g \sum_{k, k'} \psi_{\uparrow}^{\dagger}(k) \psi_{\downarrow}^{\dagger}(-k) \psi_{\downarrow} (-k') \psi_{\uparrow} (k'),
\end{equation}
where the bare coupling constant $g$ is negative, to express the attractive ({\it s}-wave) interaction between the spin-$\uparrow$ and -$\downarrow$ fermionic atoms.

After the mean-field (MF) approximation and the diagonalization of the expression for $H_{MF}$, we arrive at the following expression for the thermodynamic potential, from which all thermodynamical quantities of interest can be obtained:

\begin{eqnarray}
\label{tp}
\Omega&=&-\frac{\Delta^2}{g} \\
\nonumber
&+&  \sum_{k} \Big[ \epsilon_k^{+}-E_k-T \ln(e^{-\beta {\cal{E}}_k^{a}}+1)-T \ln(e^{-\beta {\cal{E}}_k^{b}}+1)\Big],
\end{eqnarray}
where, for simplicity of notation we have labeled $\downarrow=a$, $\uparrow=b$. Here $\beta=1/T$, where we have set the Boltzmann constant equal to $1$. We have also defined ${\cal{E}}_k^{a,b}= E_k \pm \epsilon_k^{-}$ as the quasiparticle excitations, with $E_k=\sqrt{ {\epsilon_k^{+}}^2+\Delta^2 }$, $\epsilon_k^{\pm} = \frac {\epsilon_k^a \pm \epsilon_k^b}{2}$, and the constant pairing gap is given by $\Delta = - g \int \frac{d^2 k}{(2\pi)^2} \langle \psi_{a}^{\dagger}(-k) \psi_{b}^{\dagger}(k) \rangle = \Delta^*$.

To regulate the ultraviolet divergence associated with the zero temperature term in Eq.~(\ref{tp}), we introduce the 2D bound-state equation~\cite{Randeria}

\begin{equation}
\label{reg} 
-\frac{1}{g}= \int \frac{d^2 k}{(2 \pi)^2} \frac{1}{2\xi_k+|\epsilon_B|},
\end{equation}
where $\epsilon_B$ is the 2D two-body binding energy. In addition to the regularization of the ultraviolet divergence present in Eq.~(\ref{tp}), and consequently in the gap equation (see below in Sec.~\ref{gmb}), this equation relates the strength $g$ of the contact interaction with $\epsilon_B$, which is more physically relevant, as will be clear now. In order to make contact with current experiments, it is convenient to relate $\epsilon_B$ to the three dimensional scattering length $a_s$. In the scattering of atoms confined in the axial direction by a harmonic potential with characteristic frequency $\omega_L$ they are related by~\cite{Petrov,Tempere}

\begin{equation}
\label{EB}
|\epsilon_B|=\frac{C \omega_L}{\pi} exp\left( \sqrt{2 \pi} \frac{l_L}{a_s} \right),
\end{equation}
where $a_s$ is the 3D {\it s}-wave scattering length, $\omega_L=\sqrt{8 \pi^2 V_0/ (M \lambda^2)}$, $l_L=1/\sqrt{M\omega_L}$, and $C \approx 0.915$. $V_0$ is the amplitude of the periodic potential $V_0 \sin^2(2 \pi z/ \lambda)$ generated by two counter-propagating laser beans with length $\lambda$ parallel to the $z$ axis~\cite{Tempere}.

\section{Basic Mean-Field Zero Temperature Results of 2D Imbalanced Fermi Systems}
\label{BMF}

In this section we borrow from Refs.~\cite{Randeria,Loktev,T0} some basic zero-temperature MF results, which will be needed in the next section. 

\subsection{Balanced systems}

The gap and number equations are obtained by $\partial \Omega/ \partial \Delta =0$ and $n_{\alpha}=- \partial \Omega / \partial \mu_{\alpha}$, respectively. For the balanced system where $h=0$, we find

\begin{equation}
\sqrt{\mu^2+\Delta^2}-\mu = |\epsilon_B|,
\label{gap}
\end{equation}
and
\begin{equation}
\sqrt{\mu^2+\Delta^2}+ \mu = 2 \epsilon_F,
\label{mu}
\end{equation}
where the two-dimensional Fermi energy is defined as $\epsilon_F = \frac{\pi n_T}{M}$, with $ n_T=n_a+n_b$. In the balanced configuration $n_{a}=n_{b} \equiv n = \frac{n_T}{2}$. Solving these two equations self-consistently we arrive at the well-known results~\cite{Randeria,Loktev}

\begin{equation}
\Delta_0=\sqrt{2\epsilon_F | \epsilon_B |},
\label{gap0}
\end{equation}
and

\begin{equation}
\label{mu0}
\mu_0=\epsilon_F - \frac{|\epsilon_B|}{2}.
\end{equation}

The value of the free energy at the minimum is

\begin{equation}
\Omega(h=0, \Delta=\Delta_0) \equiv \Omega_0 = - \kappa \left(\mu_0 + \frac{|\epsilon_B|}{2} \right)^2,
\label{min1}
\end{equation}
where $\kappa \equiv \frac{M}{2 \pi}$, whereas the energy of the balanced normal state is given by

\begin{equation}
\Omega(h=\Delta=0) \equiv \Omega_{bal}^N = - \kappa \mu_0^2.
\label{min2}
\end{equation}
A direct comparison between Eqs.~(\ref{min1}) and (\ref{min2}) shows that the superfluid state is energetically preferable to the normal state for any $\epsilon_B \neq 0$. Since a two-body bound state exists even for an arbitrarily small attraction in 2D~\cite{Randeria}, the pairing instability will always happens in 2D balanced two-component Fermi systems at $T=0$.

\subsection{Imbalanced systems}

We now turn our attention to the cases where $h \neq 0$. The free energy of the imbalanced normal state, $\Omega(h,\Delta=0) \equiv \Omega_{imb}^N$, is given by

\begin{eqnarray}
\Omega_{imb}^N= \int_{0}^{k_F^a} \frac{d^2 k}{(2\pi)^2} \epsilon_k^a + \int_{0}^{k_F^b} \frac{d^2 k}{(2\pi)^2} \epsilon_k^b,
\end{eqnarray}
which gives, after the integration in $k$, the free energy of a (normal) two species gas of fermionic atoms in two dimensions

\begin{eqnarray}
\label{fn}
\Omega_{imb}^N (\mu_a, \mu_b) &=& -\frac{\kappa}{2} [(\mu-h)^2 + (\mu+h)^2]
\nonumber
\\
&=& -\frac{\kappa}{2}[\mu_a^2 + \mu_b^2],
\end{eqnarray}
where we are considering $\mu_{\alpha}$ and $h$ positive.

From the graphical analysis of $\Omega$ as a function of $\Delta$ for various asymmetries~\cite{Caldas1,Caldas2,T0}, one sees that increasing the imbalance $h$ and keeping $\mu$ fixed, the minimum is still located at $\Delta_0$ up to a maximum or critical imbalance $h_c$, after which there is a quantum phase transition to the normal state with $\Delta=0$. $h_c$ is found through the equality $\Omega_0=\Omega_{imb}^N$, which yields

\begin{equation}
\label{clog}
h_c^2= \mu_0 |\epsilon_B| + \left( \frac{\epsilon_B}{2} \right)^2.
\end{equation}
From this one easily finds, by plugging into Eq.~\ref{clog} $\mu_0$ from Eq.~(\ref{mu0}) in the BCS limit ($|\epsilon_B| << \epsilon_F$): $h_c=\frac{\Delta_0}{\sqrt{2}}$, which is the same ${\rm 3D}$ result known as the Chandrasekhar-Clogston limit of superfluidity~\cite{Chandrasekhar,Clogston}.

\section{The Phase Diagram Beyond The Mean Field}
\label{GMB}

Now we construct the phase diagram of the model at finite temperature considering corrections beyond the mean field, taking into account the GMB correction.

\subsection{Induced interaction in a spin polarized Fermi gas}
\label{Ind}

The induced interaction was obtained originally by GMB in the BCS limit by the second-order perturbation \cite{Pethick00}. For a scattering process with $p_1+p_2\rightarrow p_3+p_4$, the induced interaction for the diagram in Fig.~1 is expressed as

\begin{equation}
U_{\mathrm{ind}}( p_1, p_4)= -g^2\, \chi_{ph}(p_1-p_4),
\end{equation}
where $p_{i}=({\bf k}_{i}, \omega_{l_i})$ is a vector in the space of wave-vector ${\bf k}$ and fermion Matsubara frequency $\omega_{l}=(2l+1)\pi/(\beta)$. Including the induced interaction, the effective pairing interaction between atoms with different spins is given by

\begin{figure}[htb]
  \vspace{0.5cm}
\epsfysize=5.0cm
  \epsfig{figure=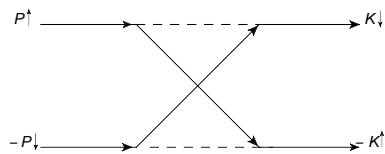,angle=0,width=5cm}
  \label{esc}
\caption[]{\label{omega} The lowest-order diagram representing the induced interaction $U_{\mathrm{ind}}( p_1, p_4)$. Arrowed and dashed lines describe fermionic propagators and the coupling $g$ between the fermionic atoms.}
\end{figure}

\begin{eqnarray}
U_{eff}( p_1, p_4)\equiv U_{eff}&=&
g +U_{\mathrm{ind}}( p_1, p_4).
\label{induce}
\end{eqnarray}
The polarization function $\chi_{ph}(p')$ is given by

\begin{eqnarray}
\chi_{ph}(p')&=& {1\over \hbar^2 \beta {\rm A}}\sum_p
\mathcal{G}_{0 b}(p)\mathcal{G}_{0 a}(p+p')\\
\nonumber
&=&\int {\rm{d}^2{\bf k}\over (2\pi)^2} {f_{{\bf k}}^b-f_{{\vec k}+{\vec q}}^a\over i
\Omega_l+\epsilon_{{\vec k}}^a-\epsilon_{{\vec k}+{\vec q}}^b},
\label{chi0}
\end{eqnarray}
where $p'=(\vec{q}, \Omega_{l})$, $\Omega_{l}=2l\pi/\beta$ is the Matsubara frequency of a boson, ${\rm A}$ is the area of the system, and $f(k)$ is the Fermi distribution function $f({{\cal{E}}_k^{a,b}})=1/(e^{\beta {\cal{E}}_k^{a,b}}+1)$. The Matsubara Green's function of a non-interacting Fermi gas is given by $\mathcal{G}_{0\sigma}(p)=1/(i\omega_l-\epsilon_{\bf k\sigma})$.  The static polarization function is then

\begin{eqnarray}
\label{chiph-10}
\chi_{ph}(|\vec q|)
&=& -2m \int {\rm{d}^2{\bf k}\over (2\pi)^2}  \frac{f_{k}^{a}}{q^2 -4mh+2kq \cos \theta} \\
\nonumber
&+&  \frac{f_{k}^{b}}{q^2 +4mh-2kq \cos \theta },
\end{eqnarray}
where $q \equiv |\vec q|$. At zero temperature the real part of Eq.~(\ref{chiph-10}) is given by

\begin{eqnarray}
\label{chiph-11-1}
\chi_{ph}(q)=\\
\nonumber
-N(0)  \left[ \frac{1}{2}- \frac{1} { 2}\sqrt{\left(1-\frac{4mh}{q^2}\right)^2-\left(\frac{2k_F^{a}}{q}\right)^2}   \right]\\
\nonumber
- N(0)  \left[ \frac{1}{2}- \frac{1} { 2}\sqrt{\left(1+\frac{4mh}{q^2}\right)^2-\left(\frac{2k_F^{b}}{q}\right)^2}   \right],\\
\nonumber
{\rm for} ~ q > k_F^{a}+k_F^{b}\\
\nonumber
=- N(0), {\rm for} ~ q \leq k_F^{a}+k_F^{b},
\end{eqnarray}
where $N(0)=\frac{m} {2 \pi }$ is the 2D density of states at the Fermi level.

In the scattering process the conservation of total momentum implies that $\vec{k}_1+\vec{k}_2=\vec{k}_3+\vec{k}_4$, with $\vec{k}_1=- \vec{k}_2$ and $\vec{k}_3=-\vec{k}_4$. $q$ is equal to the magnitude of $\vec{k}_1+\vec{k}_3$, so $q=\sqrt{(\vec{k}_1+\vec{k}_3).(\vec{k}_1+\vec{k}_3)}=\sqrt{\vec{k}_1^2+\vec{k}_3^2+ 2\vec{k}_1. \vec{k}_3}=\sqrt{\vec{k}_1^2+\vec{k}_3^2+ 2|\vec{k}_1|| \vec{k}_3|\cos \phi}$, where $\phi$ is the angle between $\vec{k}_1$ and $\vec{k}_3$. Since both particles are at the Fermi surface, $|\vec{k}_1|=|\vec{k}_3|=k_F=\sqrt{2M\mu}$, thus, $q=k_F\sqrt{2(1+\cos \phi)}$. 

The spin polarization is defined as $P=\frac{n_b-n_a}{n_b+n_a}$. Then, from Eq.~(\ref{fn}) we find $n_{\alpha}=(M/2\pi) \mu_{\alpha}$, yielding

\begin{equation}
P=\frac{h}{\mu}. 
\label{sp}
\end{equation}
We are considering $0 \leq n_a \leq n_b$, which gives $0\leq P \leq 1$. This is translated into having $0 \leq h \leq \mu$. Equation~(\ref{sp}) will enable us to express the polarization function $\chi_{ph}(h)$ of an imbalanced Fermi gas as a function of its spin polarization $P$:

\begin{eqnarray}
\label{chiph-11-2}
\chi_{ph}(q)=- N(0) \left[1- \Theta_P ~ \sqrt{1+\left(\frac{P}{\gamma}\right)^2-\frac{2}{\gamma}}\right],
\end{eqnarray}
where $\gamma=1+\cos \phi$ and $\Theta_P\equiv \Theta (\cos \phi-\sqrt{1-P^2})$.

The {\it s}-wave part of the effective interaction is approximated by averaging the polarization function $\chi_{ph}(q)$, which means an average of the angle $\phi$~\cite{GMB61,Pethick00,Yu09,Yu10}:

\begin{equation}
\langle\chi_{ph}(q)\rangle = \frac{1}{2 \pi} \int_{0}^{2 \pi} d  \phi ~\chi_{ph}(q) \equiv \chi_{ph}(h).
\end{equation}

\subsection{The GMB correction to the mean-field transition temperature and tricritical point}
\label{gmb}

Now we use the induced interaction effects calculated in Sec.~\ref{Ind} in order to obtain the corrected beyond-mean-field transition temperature of a balanced Fermi gas, and the tricritical point, $P_{tc}=(h_{tc},T_{tc})$, of an imbalanced Fermi gas.

Extremizing the grand potential of Eq.~(\ref{tp}) with respect to $\Delta$ and passing to integrals, we obtain a gap equation. The critical temperature $T_c$ is, by definition, the temperature at which $\Delta=0$. Then we find:

\begin{eqnarray}
\label{ap9}
&-&\frac{1}{g}- \chi_{ph}(h) \\
\nonumber
&-&\int \frac{d^2 k}{(2 \pi)^2}  \frac{1}{2 \epsilon_k} \left[ 1-f({\epsilon_k^{a}}) - f({\epsilon_k^{b}}) \right] \\
\nonumber
&=&\int \frac{d^2 k}{(2 \pi)^2} \frac{1}{2\xi_k+|\epsilon_B|} - \chi_{ph}(h)\\
\nonumber
&-&\int \frac{d^2 k}{(2 \pi)^2}  \frac{1}{2 \epsilon_k} \left[ 1-f({\epsilon_k^{a}}) - f({\epsilon_k^{b}}) \right] =0,
\end{eqnarray}
where $\epsilon_k=k^2/2m-\mu$. We have used Eq.~(\ref{reg}) to regulate the ultraviolet divergence, associated with the zero temperature logarithmically divergent term in Eq.~(\ref{ap9}).
The difference between the equation above and the usual mean-field thermal gap equation is that the particle-hole fluctuation has been taken into account through the effective {\it s}-wave interaction $U_{eff}$. 

Thus, to find the transition temperature of a balanced Fermi gas below which the formation of Cooper pairs becomes favorable, we have gone beyond the simple BCS approach and considered the second-order interaction effects calculated in Eq.~(\ref{chiph-11-1}). The fact that both scattering particles are at the Fermi surface restricts the angle between $\vec{k}_1$ and $\vec{k}_3$ to the single value $\phi=0$, yielding $q=2k_F$. This gives $\chi_{ph}(h=0)=-N(0)$, which results in

\begin{equation}
T_{c,GMB}=\frac{T_{c,MF}}{e} \approx \frac{T_{c,MF}}{2.72},
\label{TcGMB}
\end{equation}
agreeing completely with a previous work performed in a slightly different manner, but also considering the GMB correction~\cite{Baranov}. The result above shows that the suppression of the transition temperature by second order interaction effects is higher in 2D than in 3D, where $T_{c,GMB}^{3D} \approx \frac{T_{c,MF}^{3D}}{2.22}$~\cite{GMB61,Pethick00,Baranov2,Yu09}.

In oder to obtain the GMB correction to the mean-field tricritical point, we define from Eq.~(\ref{ap9})

\begin{eqnarray}
\label{ap10}
f(T,h)&=&\int \frac{d^2 k}{(2 \pi)^2} \frac{1}{2\xi_k+|\epsilon_B|} - \chi_{ph}(h)\\
\nonumber
&-& \int \frac{d^2 k}{(2 \pi)^2}  \frac{1}{4 \epsilon_k} \left[ \tanh({\frac{\beta \epsilon_k^{a}}{2}}) + \tanh({\frac{\beta \epsilon_k^{b}}{2}}) \right].
\end{eqnarray}
A graphical inspection shows that at the (tri)critical chemical potential imbalance $h_{tc}$, both $f(T,h)$ and its derivative with respect to $T$, $g(T,h) \equiv d f(T,h)/dT$, are zero~\cite{JSTAT1}. The vanishing of these two functions corresponds to $\alpha=\beta=0$. $\alpha$ and $\beta$ are the first and second coefficients of the expansion of the free energy in terms of the gap parameter, according to the Landau theory of phase transitions. By this criterion, one finds $h_{tc}$ and $T_{tc}$ of the tricritical point. Below this point, i.e., at low temperatures, the transition is of first order and the critical temperature has to be found by properly equating the energies of the normal and superfluid phases. Thus, for a given $h_{tc}<h<h_{c}$, where $h_c$ is the Chandrasekhar-Clogston limit of superfluidity mentioned just below Eq.~(\ref{clog}), the temperature where $\Omega(\Delta=0,T)=\Omega(\Delta_0(T),T)$ corresponds to the first order phase transition critical temperature.

\begin{figure}[htb]
  \vspace{0.5cm}
\epsfysize=6.0cm
  \epsfig{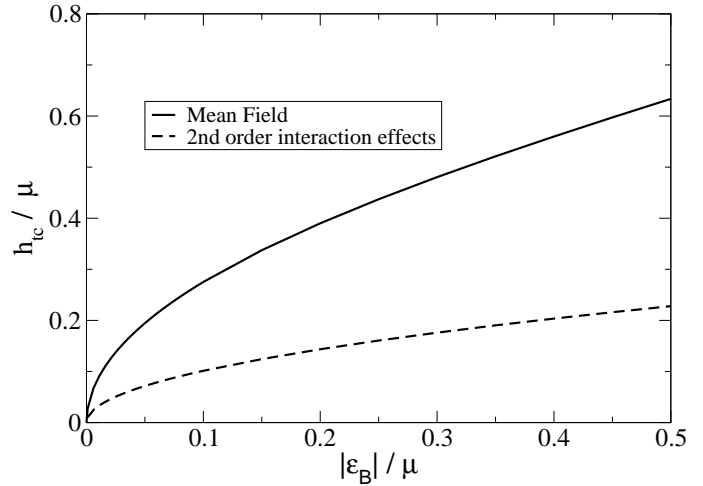}
\caption[]{\label{omega} The critical polarization $P_{c}=\frac{h_{tc}}{\mu}$ of the tricritical point plotted as a function of the two-body binding energy. The solid line shows the MF result, and the dashed line shows the results corrected by second-order interaction effects.}
\label{P}
\end{figure}

\begin{figure}[htb]
  \vspace{0.5cm}
\epsfysize=6.0cm
  \epsfig{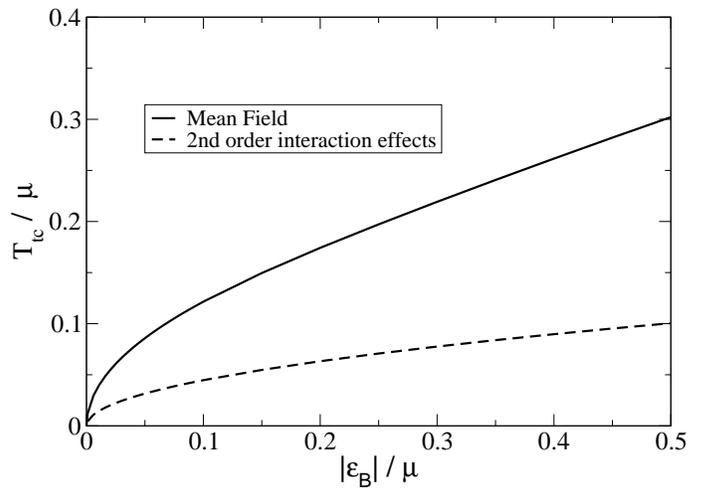}
\caption[]{\label{omega} The tricritical temperature $T_{tc}$ plotted as a function of $|\epsilon_B|/\mu$. The solid line shows the MF result, and the dashed line shows its corrections by second-order interaction effects.}
\label{T}
\end{figure}

Our results for the induced interaction corrections to the mean-field tricritical polarization and temperature are shown in Figs.~\ref{P} and~\ref{T}, respectively. As one can see in these figures, the second-order interaction effect in the BCS regime reduces the tricritical point by approximately the same amount found analytically for the balanced scenario. It is easy to verify that in the BEC region, the effect of the induced interaction vanishes due to disappearance of the Fermi surface. From Eq.~(\ref{chiph-11-1}) we obtain that $\chi_{ph}(q)$ goes to zero in the limit $k_F^a=k_F^b=0$.

\section{Conclusions}
\label{Conc}

We have studied the effects of the induced interactions on the transition temperature of a quasi-2D imbalanced Fermi gas of atoms in the BCS region. We find that the transition temperature is reduced by a factor $\approx 2.72$, in the case of a balanced Fermi system, and the temperature and polarization of the tricritical point of imbalanced Fermi systems, are also suppressed by approximately the same factor as a function of $|\epsilon_B|/\mu$, in comparison to the MF results.

It is worth mentioning that in the BEC region the universal behavior found for the transition temperature in 2D fermionic systems, considering the phase fluctuation effects~\cite{Sa,Duan2,Devreese}, is that the transition temperature obeys a behavior which may be expressed qualitatively in the form $T_c/\epsilon_F \approx {\rm const} \times \tanh(|\epsilon_B|/\epsilon_F)$, reaching a limiting value for large $|\epsilon_B|/\epsilon_F$. The value of the constant is $0.075$ in Ref.~\cite{Duan2}, and $0.125$ in Refs.~\cite{Sa} and~\cite{Devreese}. Monte Carlo simulations give $T_c/\epsilon_F \approx 0.1$ for $|\epsilon_B|/\epsilon_F=10$~\cite{MC}. This behavior can be explained with very simple arguments. The phase fluctuation treatment correctly captures the physics that in the BEC side (large values of $|\epsilon_B|$) the system behaves as weakly interacting dimers of fermionic atoms, such that the transition temperature turns out to be very insensitive to the intensity of the interaction between these composite bosons. 

We remark that to study the whole BCS-BEC crossover in a strictly 2D system, more detailed calculations are required to take into account the effects of both second order interactions and phase fluctuations. 

In conclusion, taking into account the induced interactions, we have found the tricritical temperature and polarization of the superfluid phase transition of two-component quasi-2D imbalanced Fermi gases. Our results are promising for achieving this transition in the regime of BCS pairing in current experiments.

\section{Acknowledgments}

We thank Drs. M. Continentino and L. H. C. M. Nunes for enlightening discussions. H. C. and A. L. M. are partially supported by CNPq. The authors also acknowledge partial support from FAPEMIG.

\end{document}